\documentclass[cup6b]{cupbook}
\usepackage{graphicx}
\usepackage{natbib}
\usepackage{amssymb}

\title[Rome, Italy, 27--30 April 2009]
      {The coming of age of X-ray polarimetry}
\author{}
\date{}

\begin{document}
\pagenumbering{arabic}


\author[Juri Poutanen]{Juri Poutanen \\ (Astronomy Division, Department of Physics, University of Oulu, Finland)}
\chapter{Polarization properties of X-ray millisecond pulsars}

\abstract{Radiation of X-ray bursts and of accretion shocks in weakly magnetized neutron stars in low-mass X-ray binaries is produced in plane-parallel atmospheres dominated by electron scattering. We first discuss polarization produced by single (non-magnetic) Compton scattering, in particular the depolarizing effect of high electron temperature,  and then the polarization due to multiply electron scattering in a slab. We further predict the X-ray pulse profiles and polarization properties of nuclear- and accretion-powered millisecond pulsars. We introduce  a relativistic rotation vector model, which includes the effect of rotation of polarization plane due to  the rapid motion of the hot spot as well as the light bending. Future observations of the X-ray polarization will provide a valuable tool to test the geometry of the emission region in pulsars and its physical characteristics.}

\section{Introduction}

Polarization has proven to be a valuable tool in determining the geometry of the emission region in radio pulsars \cite{bla91}. For the X-ray pulsars, the data are not yet available, but their interpretation in any case is not going to be easy, because of the strong magnetic field effects on the radiation transport. Discovery of millisecond coherent pulsations during X-ray bursts in nearly 20 low-mass X-ray binaries (so called nuclear-powered millisecond pulsars, NMSP, see \cite{SB06}) and in the persistent emission of at least eight  sources (accretion-powered millisecond pulsars, AMSP, see \cite{W06,P06,P08}) opens a completely new range of possibilities. The emission in these cases is produced at the surface of a rapidly spinning, weakly magnetized neutron star. Thus the magnetic field does not affect the radiation transport  and much more reliable predictions for the radiation pattern from the surface can be obtained. The observed pulse profiles and  polarization are affected not only by general, but also special relativistic effects \cite{PG03,VP04,PB06}.  The pulse profile alone, however, does not allow to determine uniquely the pulsar geometry, while the phase dependence of the polarization angle (PA) is a powerful tool to distinguish between the models and to put strong constraints on the geometry. In this review, we first discuss the physics of polarization produced by Compton scattering.  Then we apply these results to predict  the polarization properties of NMSP and AMSP.

\section{Polarization properties of Compton scattering}

\subsection{Polarization in single Compton scattering}
\label{sec:pol_Te}

\begin{figure}[h]
\centering
\includegraphics[scale=0.45]{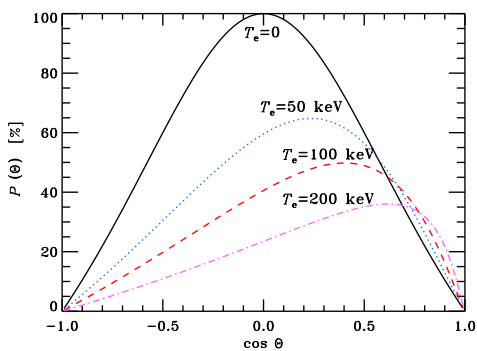}
\caption{PD at 1 keV of 0.1 keV blackbody photons scattered by the hot isotropic electrons of various temperatures $T_{\rm e}$ as a function of the scattering angle (adapted from \cite{Pou94ApJS}). For the dependence of the PD on photon energy and for cases of initially polarized radiation, see \cite{NP94}. 
}
\label{fig:pol_Te}
\end{figure}

Radiation Thomson scattered once by cold electrons becomes linearly polarized with the PD depending strongly on the scattering angle $\Theta$: $P(\Theta)=\sin^2\Theta/(1+\cos^2\Theta)$. With increasing electron temperature the PD drops because of random aberrations and corresponding random rotations of the polarization plane due to the thermal motions of the electrons \cite{NP94,Pou94ApJS} (see Fig. \ref{fig:pol_Te}). For isotropic relativistic electrons of arbitrary energies, the exact analytical expressions for the scattering redistribution matrix for Stokes parameters was derived in \cite{NP93}. For $T_{\rm e}=50$ keV, polarization is smaller by 50\% and for 100 keV it drops by a factor of two compared to the  Thomson scattering case. As electrons in accreting black holes and neutron stars often reach such temperatures, the depolarization effect has to be accounted for. For isotropic power-law distribution of relativistic electrons, the net scattered polarization is zero if the incident polarization is unpolarized \cite{BCS70,NP93}.   

\subsection{Scattering in optically thick atmosphere}
\label{sec:opt_thick}

In a classical problem of radiative transfer in a plane-parallel semi-infinite atmosphere with opacity dominated by Thomson scattering \cite{Cha60,Sob63}, 
the PD reaches the maximum of $\approx11.7\,\%$ at $\mu = 0$ (where $\mu$ is the cosine of the angle from the normal) and decreases to zero at $\mu = 1$ due to the symmetry: 
\begin{equation}\label{eq:polburst} 
P \approx -\frac{1-\mu}{1+3.582\mu} 11.71\% .
\end{equation}
The dominant direction of the electric vector oscillations is parallel to the slab plane.

\subsection{Comptonization in optically thin atmosphere}
\label{sec:opt_thin}

\begin{figure}[ht]
\centering
\includegraphics[scale=0.62]{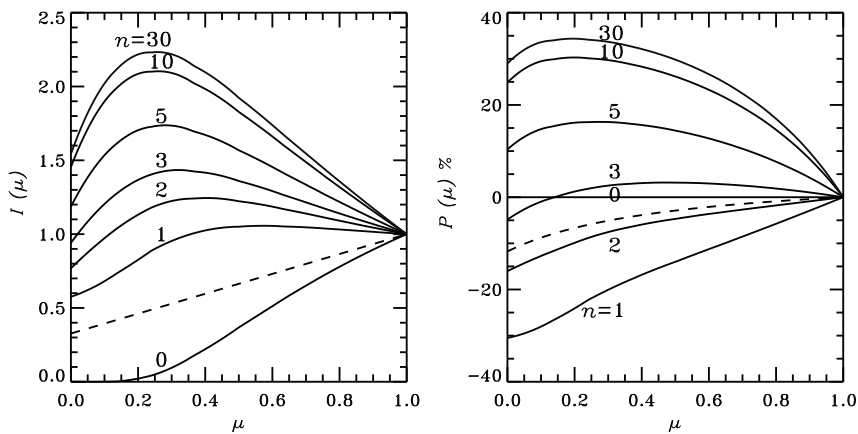}
\caption{Angular dependence of the intensity and polarization of the radiation escaping from a slab of  $\tau_{\rm T}=1$ for different scattering orders $n$. Positive $P$ corresponds to the plane of polarization parallel to the slab normal.  The unpolarized seed photons with $I(\mu)=$const were injected from the slab bottom. The dashed curves  correspond to the classical results of Chandrasekhar-Sobolev for $\tau_{\rm T}=\infty$. 
From \cite{VP04}. }
\label{fig:slab_pol}
\end{figure}

The detailed study of the transport of polarized radiation in plane-parallel atmospheres with absorption  and Thomson scattering included was considered in \cite{LS81,LS82}. The absorption causes the rotation of the polarization plane by 90$^{\rm o}$ (Nagirner effect, \cite{nag62}) even in optically thick atmospheres. The same happens when the scattering optical depth becomes   small $\tau_{\rm T}\lesssim 1$ \cite{ST85,HM93,VP04}. We are often interested in polarization of photons undergoing a certain number of scattering (because photons gain energy every time they are scattered by hot electrons). The dependence of intensity and PD on angle relative to the normal for various scattering orders is shown in Fig. \ref{fig:slab_pol} for $\tau_{\rm T}=1$. 

\begin{figure}[ht]
\centering
\includegraphics[scale=0.66]{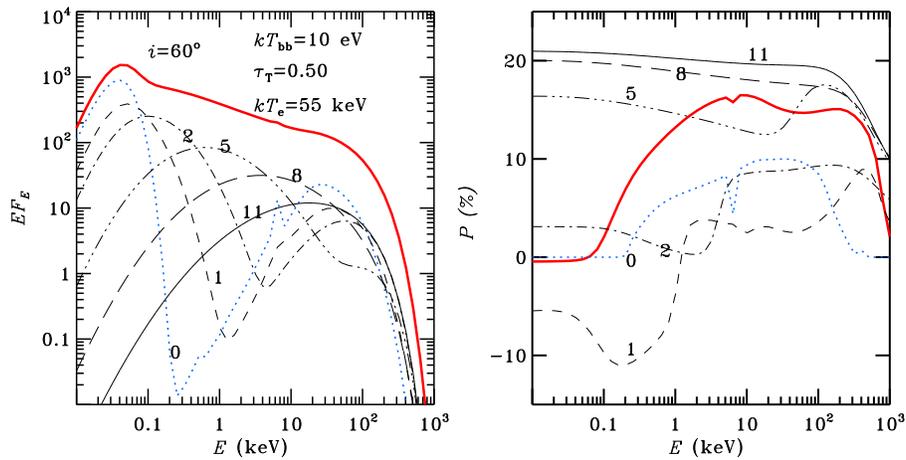}
\caption{The flux and the PD emergent from  a slab of $\tau_{\rm T}=0.5$ and electron temperature $kT_{\rm e}=55$ keV at inclination of 60$^{\rm o}$. The incident photons are blackbody of temperature 10 eV. Different scattering orders are marked at corresponding curves. The thick solid curves correspond to the total radiation. The zeroth order radiation consists of unscattered blackbody and a polarized Compton reflection component at high energies (computed using Green's matrix from \cite{PNS96}).
Adapted from \cite{PS96}. }
\label{fig:slabcorona}
\end{figure}

Usually polarization was considered in  Thomson scattering approximation, while, as we have seen in \S\ref{sec:pol_Te}, the polarization does depend on the electron temperature. Calculations of the PD in a slab using fully relativistic kernel from \cite{NP93} is shown in Fig. \ref{fig:slabcorona}.   One sees that radiation scattered once is polarized perpendicular to the slab normal (as for optically thick atmosphere), while at higher scattering orders polarization changes the sign.

\section{Polarization of accreting neutron stars}

\subsection{Relativistic rotating vector model}

When considering polarization of radiation from coherently oscillating sources, we can assume that the emission originates in one or two (in the case of AMSP) spots at the neutron star surface. In AMSP the hard X-rays are produced by thermal Comptonization in a plane-parallel slab (accretion shock) of optical depth of order unity \cite{PG03,P08}. The emission during the X-ray bursts is produced in a semi-infinite electron scattering dominated atmosphere. It is natural to assume that the radiation pattern has azimuthal symmetry in the comoving frame of a spot. Once we have the Stokes parameters in spot frame, we can transform them to the observer frame. First, we make the Lorentz transformation to the non-rotating frame accounting for Doppler boosting and relativistic aberration and then follow photon trajectories to the observer at infinity in Schwarzschild space-time. Deviations from the Schwarzschild  metric and from sphericity of the star due to the stellar rotation have a small effect and are usually neglected. For pulsar rotational frequencies of $\nu\gtrsim 400$ Hz, we  also need to account for time delays which in the extreme cases can reach about 5--10 per cent of the pulsar period.

\begin{figure}[h]
\centering
\includegraphics[scale=0.45]{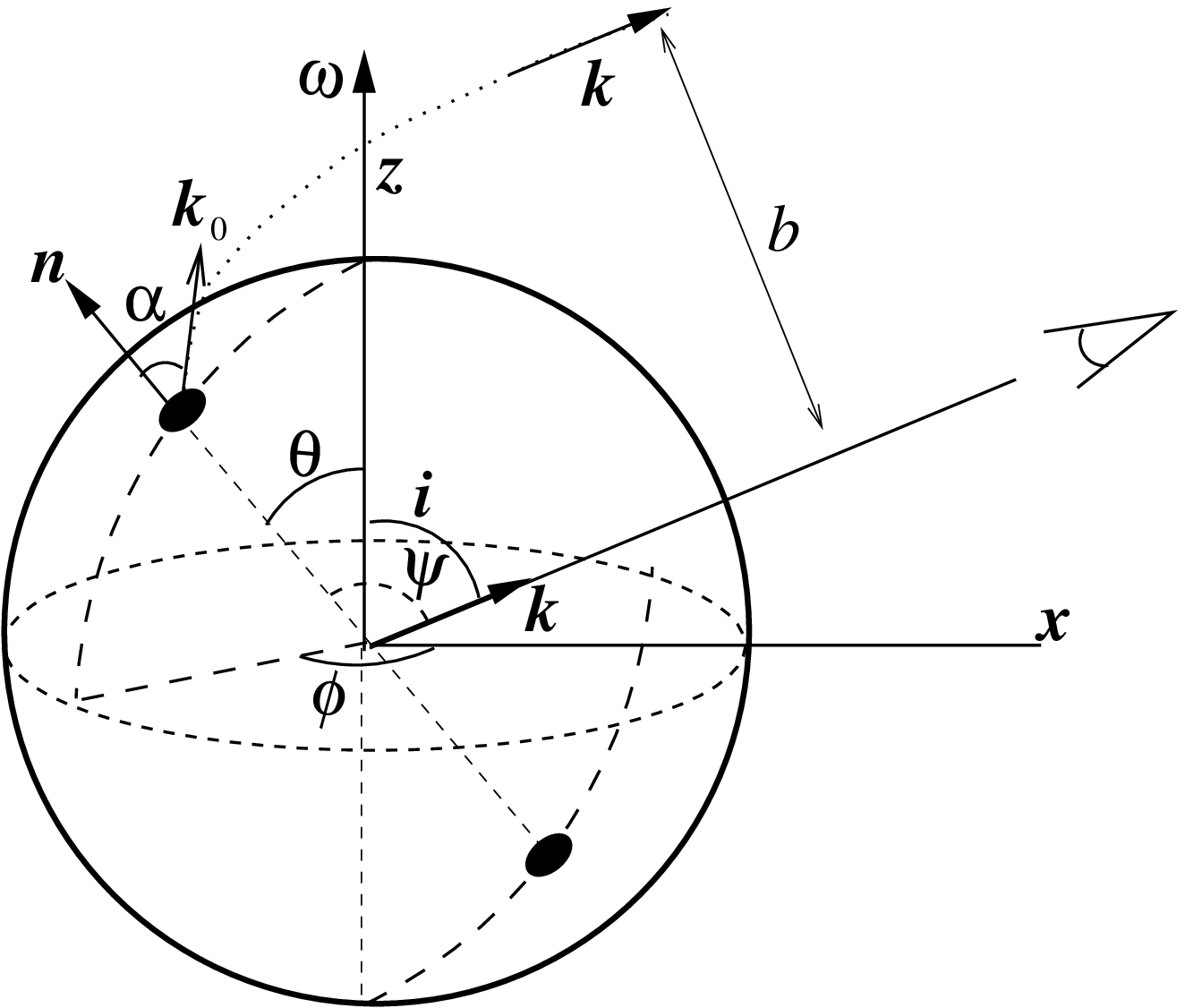}
\caption{Geometry of the emitting regions for AMSP.}
\label{fig:geom}
\end{figure}

Lorentz transformation and gravitational light bending do not change the PD and the observed value corresponds to the polar angle $\alpha'$ at which a photon is emitted in the spot comoving frame. Lorentz transformation gives $\cos\alpha'=\delta\cos\alpha$, where $\delta$ is the Doppler factor and $\alpha$ is the angle in the non-rotating frame, which is related to the position angle  $\psi$ by the usual gravitational bending formula  (see Fig. \ref{fig:geom} and \cite{PG03,VP04,PB06} for details).  

For a slowly rotating star, the polarization vector lies in the plane formed by the spot normal and the line of sight. The PA (measured from the projection of the spin axis on the plane of the sky in the counter-clockwise direction) is given by  
\begin{equation} \label{eq:rc69}
\tan\chi_0=-\frac{\sin \theta\ \sin \phi}
{\sin i\ \cos \theta  - \cos i\ \sin \theta\  \cos \phi }, 
\end{equation}
as in the rotating vector model of Radhakrishnan \& Cooke \cite{rc69}. Here $\phi$ is the pulsar phase,  $\theta$ is the spot colatitude, and $i$ is the observer's inclination (see Fig. \ref{fig:geom}). The effect of the stellar spin on rotation of the polarization vector is discussed in \cite{fer73,fer76}. Viironen \& Poutanen \cite{VP04} presented the correction to the expression for the PA which includes light bending and aberration: 
\begin{equation} \label{eq:VP04}
\tan \chi_{\rm c}=\beta_{\rm eq}  \cos\alpha  \sin\theta  \frac{\cos i\ \sin \theta -
\sin i\ \cos \theta\ \cos \phi}
{\sin \alpha\ \sin\psi +\beta_{\rm eq}  \sin\theta \sin i\ \sin\phi} ,
\end{equation}
where $\beta_{\rm eq}$ is the equatorial stellar velocity in units of speed of light. The total polarization angle for each spot is then 
$\chi =   \chi_0 + \chi_{\rm c}$.

\begin{figure}
\centering
\includegraphics[scale=0.65]{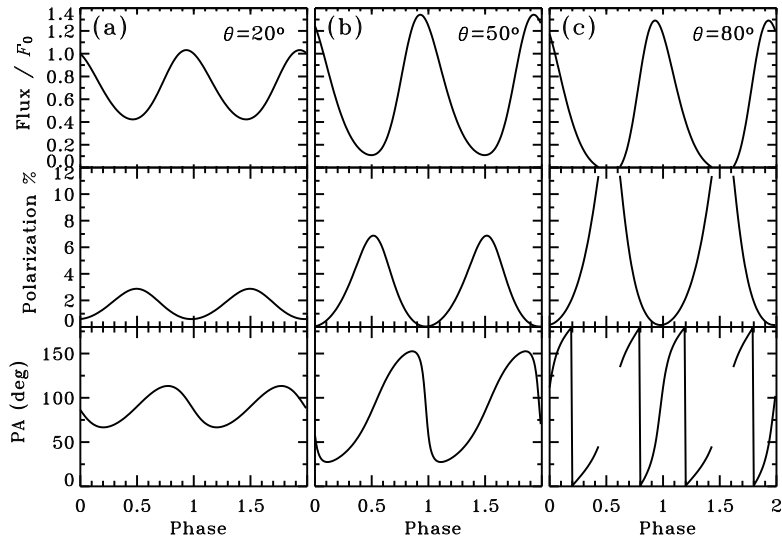}
\caption{Light curves, PD and PA expected from an X-ray burst happening at various colatitudes $\theta$. Semi-infinite electron scattering atmosphere is considered. The neutron star mass is $M=1.4{\rm M_\odot}$, radius $R=10.3$ km, rotational frequency $\nu=400$ Hz, and the observer inclination $i=60^{\rm o}$. From \cite{VP04}. }
\label{fig:pol_burst}
\end{figure}

\subsection{Polarization properties of NMSP}

The  X-ray bursts often show coherent oscillations \cite{SB06}. The energy dissipation takes place deep in the atmosphere, and thus we can assume that optical depth is infinite. At sub-Eddington luminosities the atmosphere is pinned down to the neutron star surface, and thus a plane-parallel approximation is valid. At effective temperatures of about 2 keV, most of the material is ionized and scattering dominates the opacity.  The radiation escaping from the surface then can be described by Chandrasekhar-Sobolev formulae from \S\ref{sec:opt_thick}. The predicted pulse profiles and behavior of PD and PA are shown in Fig.~\ref{fig:pol_burst} for a small bright spot.  The polarization is increasing with the spot colatitude $\theta$, reaching the maximum of $\sim 12\%$ close to the eclipses. The PA varies around $90^{\rm o}$ as the electric vector is predominantly perpendicular to the meridional plane  and its variability amplitude grows with $\theta$. The larger is the spot, the smaller is the polarization. 

\begin{figure}
\centering
\includegraphics[scale=0.6]{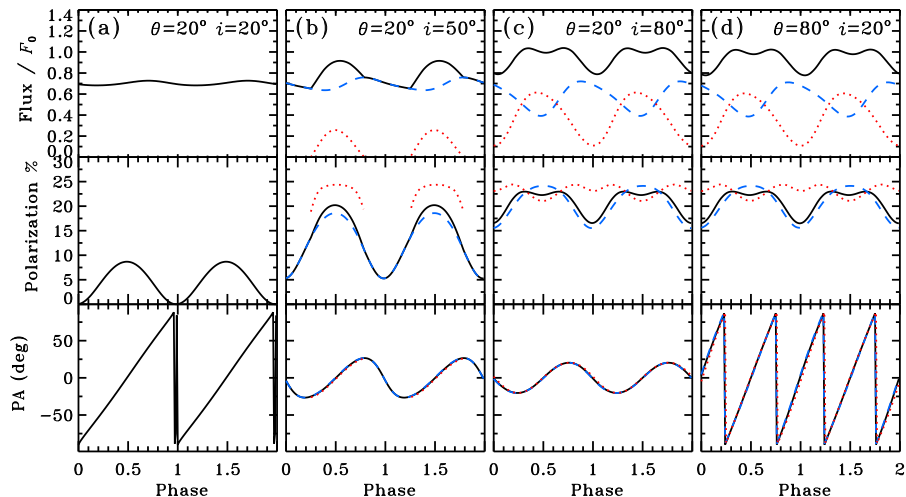}
\caption{Normalized pulse profiles, PD and PA from two antipodal point-like spots for various  inclinations and magnetic inclinations. The intrinsic radiation is scattered 7 times in a slab of Thomson optical depth $\tau_{\rm T} = 1$. Dashed curves correspond to the primary spot, dotted curves to the antipodal spot, and solid curves to the total signal.  The stellar parameters are $M=1.4{\rm M_\odot}$, $R=10.3$ km,  $\nu=300$ Hz. From \cite{VP04}. }
\label{fig:pol_amsp}
\end{figure}

\subsection{Polarization properties of accreting millisecond pulsars}

The spectra of accreting millisecond pulsars can be represented as a sum of a black body-like emission and a Comptonized tail (see review \cite{P08} and references therein). The black body photons play also a role of seed photons for Comptonization.  This interpretations is supported by the shape of the pulse profile above 7 keV which is consistent with being produced by the hotspot with the angular pattern characteristic for an optically thin slab \cite{PG03}. Because below a few keV contribution of the black body is large the expected polarization degree is small. Scattering in the hot electron slab modifies significantly the angular distribution of radiation and produces polarization signal as described in \S~\ref{sec:opt_thin}. The polarization degree is a strong function of  the scattering order for small $n$ (see Fig.~\ref{fig:slabcorona}) and therefore of the photon energy also. After a few scatterings polarization saturates. On Fig. \ref{fig:pol_amsp} we show the prediction for the polarization signal for $n=7$ (in Thomson approximation), which would correspond to about 10 keV for a typical electron temperature of 50 keV and the seed photon temperature of 0.6 keV. We choose the slab optical depth $\tau_{\rm T}=1$ which is consistent with the spectra of SAX J1808.4$-$3658 \cite{PG03,IP09}. We see that exchanging  $i$ and $\theta$ does not affect the pulse shape and PD, but the PA changes dramatically.

\section{Conclusions}

Polarization properties of non-magnetic Compton scattering, which is the main emission mechanism in weakly magnetized accreting neutron stars, have been studied in details.  Light bending in the gravitational field of the neutron star and the rotation of the polarization plane because of relativistic effects are also well understood. Thus the theory of the polarization from millisecond X-ray pulsars is rather well developed by now and is ready to be applied to the data. The launch of an X-ray polarimeter will hopefully start a new era in studies of these sources and will provide  a powerful tool for understanding the geometry and the physics of pulsars.

\end{document}